\documentstyle[preprint,prl,aps]{revtex}

\begin{document}

\draft

\preprint{\vbox{
\hbox{CTP-TAMU-48/95}
\hbox{hep-ph/9512268}
\hbox{December 1995}
}}

\title{Gravitational GUT Breaking and the GUT-Planck Hierarchy}

\author{S.~Urano\cite{shinichi}, D.~Ring\cite{dave}, and
R.~Arnowitt\cite{dick}}

\address{Center for Theoretical Physics, Department of Physics,
		Texas A\&M University,
  		College Station, TX 77843-4242}

\maketitle

\begin{abstract}

It is shown that non-renormalizable gravitational interactions in the Higgs
sector of supersymmetric grand unified theories (GUT's) can produce the
breaking of the unifying gauge group $G$ at the GUT scale $M_{\rm GUT} \sim
10^{16}$~GeV. Such a breaking offers an attractive alternative to the
traditional method where the superheavy GUT scale mass parameters are added ad
hoc into the theory.  The mechanism also offers a natural explanation for the
closeness of the GUT breaking scale to the Planck scale.  A study of the
minimal SU(5) model endowed with this mechanism is presented and shown to be
phenomenologically viable.  A second model is examined where the Higgs doublets
are kept naturally light as Goldstone modes. This latter model also achieves
breaking of $G$ at $M_{\rm GUT}$ but cannot easily satisfy the current
experimental proton decay bound.

\end{abstract}
\narrowtext

The precision LEP data has indicated a unification, with the help of
supersymmetry, of the three coupling constants of the Standard Model at a scale
of $M_{\rm GUT} \sim 10^{16}$~GeV, thus indicating the existence of a hierarchy
between the Planck scale and the grand unification scale.  This hierarchy has
allowed model builders to build various GUTs as effective field theories,
sometimes including non-renormalizable operator (NRO) terms scaled by inverse
powers of the Planck mass to account for gravitational effects.  Such models
with additional NRO terms can successfully account for lower values of
$\alpha_3(M_Z)$\cite{HallSarid,RingEtAl} should the lower energy measurements
of that parameter turn out to be correct. They can also affect various low
energy predictions such as that of $\tan\beta$ from the $m_b/m_\tau$
constraint\cite{DasguptaEtAl}. In these field theoretic approaches, the GUT
gauge group is broken and the hierarchy is achieved by the ad hoc inclusion of
$O(M_{\rm GUT})$ mass terms.

The smallness of the GUT-Planck hierarchy, on the other hand, suggests that the
true GUT may actually reside at the Planck scale, unified with gravity.
However, one would then expect all mass parameters to be $O(M_{\rm Pl})$, i.e.
a particle would either be massless or a member of a tower of Planck mass
states. The inclusion of $O(M_{\rm GUT})$ mass terms would then be difficult to
justify. In this letter, we point out that without any $O(M_{\rm GUT})$ mass
parameters, one may still achieve gauge group breaking via the NRO terms. In
addition, the spectrum of masses produced in this type of GUT breaking is
generally below the Planck scale, so that the coupling constant unification
naturally occurs at O($10^{16}$ GeV), despite the absence of mass parameters of
such size.  The specific spectrum and its parametrization is model dependent,
however, and we will examine below two models and their phenomenological
viabilities.

Within the context of string theory, gauge coupling unification below the
string scale of $M_{\rm string} \sim g_{string} \times 5 \times 10^{17}$~GeV
has been problematic\cite{DienesEtAl}. The models we consider make use of
adjoint representations to break the gauge group. We note that SU(5) and SO(10)
GUTs from Kac-Moody (KM) level 2 strings allow the existence of such adjoint
representations with the requirement that there be no mass terms in the
superpotential\cite{AldazabalEtAl}.  Furthermore, though as yet not realistic,
an explicit three generation SU(5) example of a KM level 2 model has now been
constructed\cite{ChaudhuriEtAl}. Although we do not consider here string models
explicitly, our work suggests the possible realization of such a model.

Our first model is a simple modification of the minimal SU(5) SUSY-GUT where
the mass term for the adjoint Higgs has been eliminated, and instead the
leading NRO self interaction terms have been added. The superpotential of the
symmetry breaking sector of our SU(5) model, up to $O(1/M_{\rm
Pl})$\cite{sigma5}, is
\begin{equation}
W = \frac{1}{3} \lambda_0 {\rm tr} \Sigma^3 +
    \frac{1}{4} \frac{\lambda_1}{M_{\rm Pl}} ({\rm tr}\Sigma^2)^2 +
    \frac{1}{4} \frac{\lambda_2}{M_{\rm Pl}} {\rm tr} \Sigma^4 + W_H
\label{eq_w1}
\end{equation}
where $\Sigma$ is a {\bf 24} of SU(5), $W_H$ couples $\Sigma$ to the {\bf 5}
and $\overline{\bf 5}$ Higgs, and $M_{\rm Pl} = 1/\sqrt{8 \pi G_N} \simeq 2.4
\times 10^{18}$~GeV. We might imagine the $\lambda_{1,2}$ terms to arise from
integrating out Planckian mass states. The effective potential for scalar
fields is given by $V = \sum_i | \frac{\partial W}{\partial \phi_i} |^2$, where
$\phi_i$ is the scalar component of the multiplet. Minimizing the effective
potential, one finds the VEV of $\Sigma$ that breaks the gauge group into SU(3)
$\times$ SU(2) $\times$ U(1),
\begin{equation}
\langle \Sigma \rangle = {\rm diag} (2,2,2,-3,-3) \Sigma_0,
\label{eq_SigmaVev}
\end{equation}
where
\begin{equation}
\Sigma_0 = \frac{\lambda_0}{(30 \lambda_1 + 7 \lambda_2)} M_{\rm Pl}.
\label{eq_Sigma0}
\end{equation}
Here, $\Sigma_0$ plays the role of $M_{\rm GUT}$. It is the smallness of this
VEV for reasonable values of $\lambda_{0,1,2}$ which leads to the GUT-Planck
hierarchy in this model. Expanding $\Sigma$ around this VEV, $\Sigma = \langle
\Sigma \rangle + \Sigma'$, one finds that the (${\bf 3}_{SU(3)}$, ${\bf
2}_{SU(2)}$, $(5/3)_{U(1)}$) components become Goldstone bosons giving rise to
a mass for the super-heavy vector bosons, $M_V = 5 \sqrt{2} g_5 \Sigma_0$,
where $g_5$ is the GUT gauge coupling constant.  Note that the large factor of
30 in the denominator of Eq.~(\ref{eq_Sigma0}), which is instrumental in
achieving the GUT-Planck hierarchy, is purely group theoretical in origin,
coming from $\langle {\rm tr}\Sigma^2\rangle$. The remaining components of
$\Sigma'$ , ( ({\bf 8},{\bf 1},0), ({\bf 1},{\bf 3},0), and ({\bf 1},{\bf 1},0)
), grow masses:
\begin{eqnarray}
M_{\Sigma_8} &=& 5 \lambda_0
			\frac{15 \lambda_1 + 4 \lambda_2}
			     {30 \lambda_1 + 7 \lambda_2} \Sigma_0,
			\nonumber \\
M_{\Sigma_3} &=& \frac{15}{2} \lambda_0
			\frac{10 \lambda_1 + \lambda_2}
			     {30 \lambda_1 + 7 \lambda_2} \Sigma_0,
			\nonumber \\
M_{\Sigma_1} &=& \frac{1}{2} \lambda_0
			\Sigma_0.
\end{eqnarray}

For $W_H$ we consider the simplest choice,
\begin{equation}
W_H = \lambda' H (\Sigma + 3 M') \overline{H},
\end{equation}
where $H$ and $\overline{H}$ are the Higgs {\bf 5} and $\overline{\bf 5}$.
Here, the mass parameter $M'$ is fine tuned to equal $\Sigma_0$ so as to
generate a pair of massless SU(2) doublets which break the electro-weak
symmetry\cite{Hnro}.  We look later at more natural solutions to this well
known fine tuning problem. The remaining components of $H$ and $\overline{H}$,
({\bf 3},{\bf 1},2/3), then grows a mass
\begin{equation}
M_{H_3} = 5 \lambda' \Sigma_0.
\end{equation}
Defining $M_U$ as the largest of these masses, we see that it is easy to get
$M_U \sim 10^{16}$~GeV, for example with $\lambda_0 = 0.1$, $\lambda_1 = 1$,
$\lambda_2 = 1$, $\lambda' = 1$ and $g_5 = 0.7$ one finds $M_U = M_{H_3} = 3.24
\times 10^{16}$~GeV. Note that this model has the same particle content as the
minimal SU(5) model except that the octet and the triplet of $\Sigma$ are not
degenerate when $\lambda_2 \ne 0$.  In general this non-degeneracy affects
coupling constant unification\cite{nonDeg} and can change the prediction of
$\alpha_3$, although not as strongly as the NRO which we discuss next.

To investigate gauge coupling unification, we add to our Lagrangian a term
$(c/2 M_{\rm Pl}) {\rm tr}(\Sigma F F)$\cite{Hill,ShafiWetterich}, which has a
significant effect on the matching conditions at the high
scale\cite{HallSarid,RingEtAl}. This term can arise naturally in supergravity
from the expansion of the gauge kinetic energy function $f_{\alpha \beta} =
\delta_{\alpha \beta} + \frac{1}{M_{\rm Pl}} a^i_{\alpha \beta \gamma}
\phi_i^\gamma + \cdots $ for $\phi_i^\gamma = \Sigma$.  (This term, when
$\phi_i^\gamma$ is the Polonyi field that spontaneously breaks supersymmetry,
similarly gives rise to the gaugino soft breaking mass, so it is not unnatural
to also have a term where $\phi_i^\gamma = \Sigma$.) A Monte Carlo exploration
has been done using the ``naturalness'' condition that $|c|$ and each
$|\lambda|$ lie between 0.1 and 2.  For each point in this parameter space,
values of $\alpha_3(M_Z)$, $\sin^2(\theta_W)$, and $g_5$  were determined by
requiring that the coupling constants unify at $M_U$. We run the coupling
constants using the 2-loop renormalization group equations (RGE's) in the
$\overline{\rm MS}$ scheme including the particle content of the MSSM and the
SU(5) GUT as described above\cite{Bfunc}. Each point is then checked to see if
it satisfies the proton decay bound, $\tau(p\rightarrow\bar{\nu}K) > 1 \times
10^{32}$~yr (90\% CL)\cite{pdecay} which leads to $M_{H_3} > 1.2 \times
10^{16}$~GeV\cite{AN94}. The results appear as Fig.~\ref{fig1}\cite{www}. We
find that phenomenologically acceptable values of $\alpha_3(M_Z)$ and
$\sin^2(\theta_W)$ generally require $\lambda_0$ to be small, $\lesssim 0.3$,
while $\lambda_1 \gtrsim 0.5$. The proton decay constraint does restrict the
parameter space but only marginally, requiring $\lambda' \gtrsim 0.2$.

We turn now to a model that avoids the doublet-triplet splitting problem. There
are several known ways to naturally make the Higgs doublets massless. Of these,
an attractive method is to assume that there is a global symmetry in the Higgs
sector of the GUT which when broken (effected by the breaking of the unifying
gauge group) results in the Higgs doublets becoming the Goldstone modes, and
thus
massless\cite{InoueEtAl,AnselmJohansen,Anselm,BarbieriEtAl,BerezhianiEtAl}. An
elegant realization of this idea is to embed the local SU(5) symmetry into a
larger global SU(6). We now examine this GUT modified in a minimal way, in the
same manner as our first model, so that the NRO's are given the role to break
the unifying gauge symmetry.  We consider a superpotential of the form
\begin{equation}
W = \frac{1}{3} \lambda_0 {\rm tr} \Sigma^3 +
    \frac{\lambda_1}{4 M_{\rm Pl}} ({\rm tr}\Sigma^2)^2 +
    \frac{\lambda_2}{4 M_{\rm Pl}} {\rm tr} \Sigma^4,
\label{Wsu6}
\end{equation}
where now $\Sigma$ is a {\bf 35} of SU(6) which decomposes to {\bf 24} + {\bf
5} + $\overline{\bf 5}$ + {\bf 1} under SU(5), and then to $({\bf 8}, {\bf 1},
0) + ({\bf 1}, {\bf 3}, 0) + ({\bf 3}, {\bf 2}, 5/3)_g + (\overline{\bf 3},
{\bf 2}, -5/3)_g + ({\bf 1}, {\bf 1}, 0) + ({\bf 3}, {\bf 1}, 2/3) +
(\overline{\bf 3}, {\bf 1}, -2/3) + ({\bf 1}, {\bf 2}, 1)_g + ({\bf 1}, {\bf
2}, -1)_g + ({\bf 1}, {\bf 1}, 0)$ under SU(3) $\times$ SU(2) $\times$ U(1),
where the Goldstone modes are indicated with a subscript $g$.  The effective
potential from Eq.~(\ref{Wsu6}) then yields the VEV which breaks the SU(6)
symmetry down to SU(4) $\times$ SU(2) $\times$ U(1) (and consequently the SU(5)
down to the standard model):
\begin{equation}
\langle \Sigma \rangle = {\rm diag} (1,1,1,1,-2,-2) \Sigma_0,
\label{eq_SigmaVev_6}
\end{equation}
where
\begin{equation}
\Sigma_0 = \frac{\lambda_0}{3 (4 \lambda_1 + \lambda_2)} M_{\rm Pl}.
\label{eq_Sigma0_6}
\end{equation}
The color triplet Goldstone modes give mass to the super-heavy vector bosons
via the Higgs mechanism while the SU(2) doublets automatically remain massless
and are identified as the light Higgs doublets. Thus the super-heavy spectrum
is:
\begin{eqnarray}
M_{\Sigma_8} = M_{H_3} = M_{\Sigma_1} &=&
	\frac{3}{2} \lambda_0 \Sigma_0, \nonumber \\
M_{\Sigma_3} &=&
	6 \lambda_0 \frac{ \lambda_1 }{4 \lambda_1 + \lambda_2} \Sigma_0,
	\nonumber \\
M_{\Sigma_1'} &=&
	\frac{1}{2} \lambda_0 \Sigma_0, \nonumber \\
M_V &=&
	3 \sqrt{2} g_5 \Sigma_0.
\end{eqnarray}

Monte Carlo investigation reveals that again, a sufficient GUT-Planck hierarchy
is naturally generated and coupling unification is achieved with the current
measurements of the coupling constants. However, we find that this time, the
values of $M_{H_3}$ in the region where the coupling unification is achieved is
below the proton decay bound of $1.2 \times 10^{16}$~GeV. The discrepancy
arises due to the fact that there commonly exists a splitting between
$M_{\Sigma_{3,8}}$ and $M_V$ in these gravitationally induced GUT breaking
models.  In particular, for this global SU(6) model, the ratio
$M_{\Sigma_8}/M_V = \lambda_0/{2 \sqrt{2} g_5}$ must be $\lesssim 0.15$ since
$\lambda_0$ needs to be $\lesssim 0.3$ in order to generate sufficient
GUT-Planck hierarchy, and $g_5 \sim 0.7$.  On the other hand, $M_V$ and
$M_{\Sigma_{3,8}}$ are related by $M_V^{2/3} M_{\Sigma_8}^{1/6}
M_{\Sigma_3}^{1/6} = 2.0 \times 10^{16}$~GeV\cite{mSigmaV} which becomes a
relation between $M_{H_3}$ and $M_V$ in this global SU(6) model where
$M_{\Sigma_8}$ is degenerate with $M_{H_3}$.  The requirement for the splitting
between $M_{H_3}$ and $M_V$ then forces $M_{H_3}$ below the proton decay bound
(unless $M_{\Sigma_3}/M_{\Sigma_8}$ is pushed to be unnaturally small).

It is possible to overcome this difficulty by going outside of what we above
defined as ``natural'' values of the parameters.  One possibility is to allow
values of $\lambda_{1,2}$ that are above the upper bound of 2.0 which assure
validity of the perturbation theory.  Such large values will help in creating
sufficient GUT-Planck hierarchy so that the above constraint on $\lambda_0$ and
thus $M_{\Sigma_8}/M_V$ is loosened. For example, $\lambda_0=0.627$, $\lambda_1
= 8.33$, $\lambda_2 = 6.03$, $g_5 = 0.735$, and $c=1.4$ give $M_\Sigma = 1.02
\times 10^{16}$~GeV, $M_V = 3.98 \times 10^{16}$~GeV, $M_{H_3} = 1.20 \times
10^{16}$~GeV, $\alpha_3(M_Z) = 0.125$, and $\sin^2(\theta_W)=0.2307$. Another
possibility is to allow the ratio $M_{\Sigma_3}/M_{\Sigma_8}$ to be unnaturally
small, which translates to allowing small values for $\lambda_1/\lambda_2$. For
example, $\lambda_0=0.136$, $\lambda_1 = 0.00205$, $\lambda_2 = 1.84$, $g_5 =
0.759$, and $c=1.38$ give $M_\Sigma = 5.34 \times 10^{16}$~GeV, $M_V = 1.9
\times 10^{17}$~GeV, $M_{H_3} = 1.2 \times 10^{16}$~GeV, $\alpha_3(M_Z) =
0.125$, and $\sin^2(\theta_W)=0.2307$. Yet another possibility is to arrange a
highly non-degenerate SUSY spectrum so that their low energy threshold
contribution (thus far assumed negligible) can favorably affect the predicted
values of $\alpha_3(M_Z)$ and $\sin^2(\theta_W)$. We find that in general, one
then needs to have a large ratio between the slepton masses and the gluino
mass.  If we assume gluino mass of O(100 GeV), one would then need the slepton
masses to be of O(5 TeV).

In conclusion, we have shown that GUT scale can be generated naturally from
$M_{Pl}$ when the GUT symmetry is broken by NRO interactions without having to
put in the GUT scale mass parameter by hand.  We have explicitly demonstrated
this mechanism within the context of minimal SU(5) GUT.  The resulting GUT is
shown to be phenomenologically viable.  The global SU(6) model, which gives a
natural doublet-triplet Higgs mass splitting, also generates a GUT mass scale
in this fashion.  However, such models have difficulty in satisfying the
current proton decay bound. Nevertheless, the mechanism is quite general and
applicable to many other GUTs.

This work was supported in part by the National Science Foundation under grant
number PHY-9411543.

\begin{figure}
\caption{The subregion of parameter space satisfying the naturalness
constraints, $0.1<|\lambda|, |c|<2$, and the proton decay constraint, $2.0
\times 10^{17}$~GeV~$> M_{H_3} > 1.2 \times 10^{16}$~GeV, projected onto the
$\sin^2(\theta_W)$-$\alpha_3(M_Z)$ plane.  The box corresponds to the
experimentally measured value of $\sin^2(\theta_W) = 0.23129 \pm 0.00060$ and
$\alpha_3(M_Z) = 0.118 \pm 0.007$.  The $\sin^2(\theta_W)$ width corresponds to
a 2 s.d.\ range around the current world average (M.~Woods, talk at
International Europhysics Conference on High Energy Physics, Brussels, 1995.)
The upper bound on $M_{H_3}$ represent a reasonable upper bound on the validity
of a GUT field theory (above which Planck physics, e.g.\ strings, effects would
become large).}
\label{fig1}
\end{figure}


\begin{thebibliography}{10}
\bibitem{sigma5}
We have also examined the case where $O(1/{M_{\rm Pl}}^2)$ terms have been
  added to the superpotential, and have found that they make only a small
  perturbation to our results.
\bibitem{Hnro}
It is possible to achieve doublet-triplet splitting with an NRO term instead of
  the $M'$ mass term by replacing $3 \lambda' M'$ with $\lambda_1'
  \Sigma^2/M_{\rm Pl} + \lambda_2' {\rm tr} \Sigma^2/M_{\rm Pl}$. The fine
  tuning then takes the form $\lambda'/(3 \lambda_1' + 10 \lambda_2') =
  \lambda_0/(30 \lambda_1 + 7 \lambda_2)$.
\bibitem{nonDeg}
The effect of superheavy thresholds to $\alpha_3$ can be summarized by
  considering the combination $2 \alpha_3^{-1} - 3 \alpha_2^{-1} +
  \alpha_1^{-1}$ of coupling constants at $M_Z$: $$ \alpha_3^{-1}(M_Z) =
  \frac{1}{2} [ 3 \alpha_2^{-1} - \alpha_1^{-1} ] - \frac{3}{5 \pi} \ln
  \frac{M_{H_3}}{M_Z} + \frac{3}{2 \pi} \ln \frac{M_{\Sigma_3}}{M_{\Sigma_8}}.
  $$ The last term, for our SU(5) model is $(3/2 \pi) \ln( (30 \lambda_1 + 3
  \lambda_2)/(30 \lambda_1 + 8 \lambda_2) )$. By taking the case where signs of
  $\lambda_1$ and $\lambda_2$ are different, this contribution can lower the
  acceptable range of $\alpha_3$, though its effect is in general small
  compared to that of the NRO $(c/2M_{\rm Pl}) {\rm tr} (\Sigma F F)$,
  mentioned later in the main text. In our numerical search, summarized in
  Fig.~1, we find that the lower bound for $\alpha_3$ is $\simeq 0.105$.
\bibitem{Bfunc}
The two loop RGE, appropriate beta function coefficients, as well as their
  changes under various threshold contributions may be found in
  \cite{RingEtAl}.
\bibitem{mSigmaV}
The combination $ -5 \alpha_1^{-1} + 3 \alpha_2^{-1} + 2 \alpha_3^{-1}$ leads
  to a one loop relation between $M_{\Sigma_{3,8}}$, $M_V$, and the low energy
  data, \begin{eqnarray} \lefteqn{ \frac{1}{12 \pi} \ln (
  \frac{M_{\Sigma_3}}{M_Z} ) + \frac{1}{12 \pi} \ln ( \frac{M_{\Sigma_8}}{M_Z}
  ) + \frac{2}{6 \pi} \ln ( \frac{M_V}{M_Z} ) } \nonumber \\ &=&\frac{1}{12}
  \frac{\cos(2 \theta_W)}{\alpha(M_Z)} - \frac{1}{18} \frac{1}{\alpha_3(M_Z)} -
  \frac{2}{18 \pi} \ln ( \frac{M_{\rm SUSY}}{M_Z} ). \nonumber \end{eqnarray}
  This expression is unaffected by the NRO, $(c/{2 M_{\rm Pl}}) {\rm tr}
  (\Sigma F F)$, and thus provides a strong relationship amongst the proton
  decay modes for the global SU(6) type GUTs where $M_{H_3} = M_{\Sigma_8}$.
  The value in the main text is gotten from the above equation with
  $\sin^2(\theta_W) = 0.23129$, $\alpha_3(M_Z) = 0.118$, and $M_{\rm SUSY} =
  M_Z$.
\end{thebibliography}
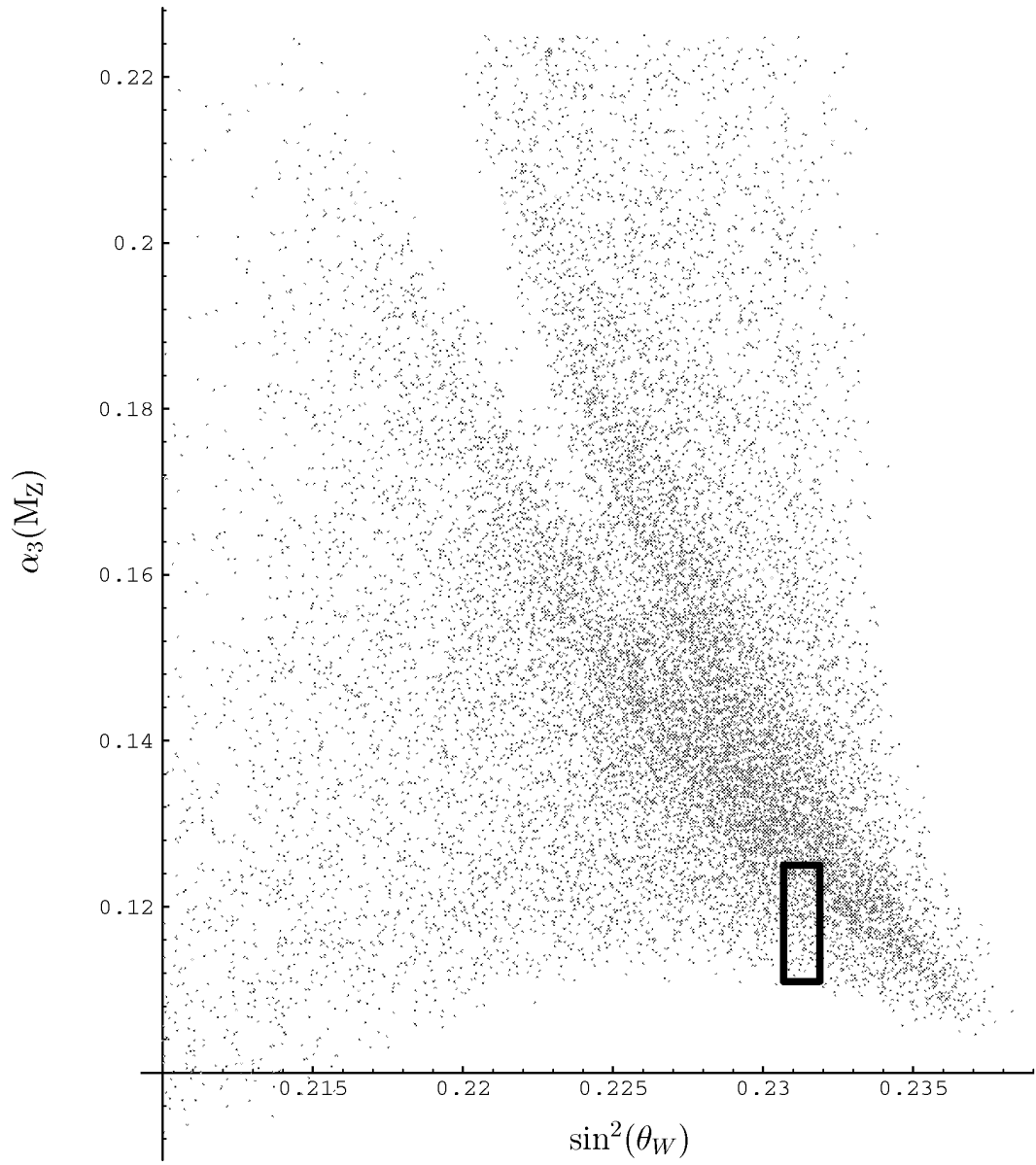
\end{document}